\documentstyle[prl,preprint,aps]{revtex}
\input{psfig}

\newcommand{\be}{\begin{equation}}
\newcommand{\ee}{\end{equation}}
\newcommand{\ba}{\begin{eqnarray}}
\newcommand{\ea}{\end{eqnarray}}

\newcommand{\nonu}{\nonumber \\[2mm]}
\newcommand{\noin}{\noindent}

\newcommand{\cd}{c^{\dagger}}

\newcommand{\la}{\langle}
\newcommand{\ra}{\rangle}

\begin{document}
\draft
\title{Ground-State Dynamical Correlation Functions: An Approach from Density
Matrix Renormalization Group Method}
\author{Hanbin Pang, H. Akhlaghpour, and M. Jarrell\\}
\address{Department of Physics, University of Cincinnati, Cincinnati, OH 45221
}
\date{\today}
\maketitle
\begin{abstract}
A numerical approach to ground-state dynamical correlation
functions from Density Matrix Renormalization Group (DMRG)
is developed. Using sum rules, moments of a dynamic correlation
function can be calculated with DMRG, and with the moments
the dynamic correlation function can be obtained by the
maximum entropy method. We apply this method to one-dimensional
spinless fermion system, which can be converted to the spin
1/2 Heisenberg model in a special case. The dynamical
density-density correlation function is obtained.
\end{abstract}

\narrowtext

Dynamical correlation functions of a model are of special interest,
because they can provide a comprehensive comparison to
experimental measurements. Unfortunately they are very difficult to calculate
analytically or numerically for strongly correlated systems. Even for
one-dimensional systems, the dynamical correlation functions are hard to
obtain. For example the $S=1/2$ Heisenberg model, although its exact
solution from Bathe {\it Ansatz} has been known for a long time, its
ground-state dynamical correlation functions have not yet been obtained.
Until now there are only a few general ways to obtain dynamical
correlation functions.
Analytically, only the asymptotic behavior of correlation functions for
one-dimensional models in the quantum critical regime are able to obtain
by bosonization or conformal field theory \cite{Luther}. Numerically,
one way to calculate
dynamical correlation functions is the analytic continuation of quantum
Monte Carlo simulations with the maximum entropy method \cite{Deisz,Preuss}.
But this method will encounter
essential difficulties if we are interested in zero-temperature
properties. Another numerical method to calculate the
ground-state dynamical properties \cite{Gagliano} is based on the
Lanczos method. The limitation of this method is that it cannot be applied
to large systems.

The Density Matrix Renormalization Group (DMRG) method
proposed by White \cite{White} is a powerful method to study the
ground state of one-dimensional interacting systems. With this method
the ground state energy, a few excitation energies, and static correlation
functions can be calculated for a large system. However, it
was not clear if one can obtain dynamical properties from this method.

In this paper, we describe a numerical method for calculating
ground-state dynamical
correlation functions in a systematic way, which is a combination of DMRG
and maximum entropy methods (MEM) \cite{Brandt}.
In general the moments of a dynamical correlation function can be
expressed as static correlation functions,
which can be calculated by DMRG method. With these moments we
can obtain the dynamic correlation function with MEM. We apply this
method to the one-dimensional spinless fermion system with nearest neighbor
interaction. This model is equivalent to the spin-1/2 $XXZ$ chain.
We have considered two special cases of this model, corresponding to
the $XY$ model and the Heisenberg model. The dynamical density-density
correlation (namely the structure function $S(q,\omega)$ in spin chain)
is obtained.
For non-interacting case (the $XY$ model) we compare our result with the exact
result, and obtain a very good agreement.

The one-dimensional spinless fermion model we consider has the following
Hamiltonian:

\be
H = -t\sum_{i} (\cd_{i}c_{i+1}+$h.c.$) + V\sum_{i} n_{i}n_{i+1},
\ee

\noin where $c^{(\dagger)}_{i}$ are annihilation (creation) operators for a
fermion at site $i$, and $n_{i}=\cd_{i}c_{i}-\frac{1}{2}$. The Hamiltonian
written in such form ensures the ground state is at half filling. This model
may be mapped to the $XXZ$ model by the Jordan-Wigner transformation. Under
this transformation $S^z_i=n_i$, $J_x=J_y=2t$, and $J_z=V$. At $V=0$ this
model is equivalent to the $XY$ model, while at $V=2t$ it
is equivalent to the Heisenberg model. We only consider these two cases in
this paper, and the results for other $V$ will be presented elsewhere
\cite{pang}.

The first step of our method is to use sum rules to express the moments of a
dynamical correlation function by some static correlation functions.
The sum rules for the spin model have been derived \cite{Hohenberg}.
We use the similar definition of
the correlation functions as in Ref. \onlinecite{Hohenberg}

\ba
\chi_{c}(q,t) &=& \frac{1}{2}\la \{n(q,t), n(-q,0)\} \ra
- \langle n(q,t) \ra \la n(-q,0) \ra,  \nonu
\chi''(q,t) &=& \frac{1}{2}\la [n(q,t), n(-q,0)] \ra ,
\ea

\noin where $n(q)=N^{-1/2}\sum n_l e^{iql}$, the curly bracket is
an anticommutator, and $\la n(q)\ra=$Tr$(n(q)e^{-\beta H})$.
The fluctuation-dissipation theorem
gives the relation: $\chi_c(q,\omega)=\coth(\omega/2k_BT)\chi''(q,\omega)$.
The structure function or dynamic form factor $S(q,\omega)$ is defined as
$S(q,\omega)=\chi''(q,\omega)/(1-e^{-\omega/k_BT})$.
Due to the parity and time reversal symmetry in our model,
$\chi''(q,\omega)$ and $\chi_c(q,\omega)$ have following
properties: $\chi''(q,-\omega)=-\chi''(q,\omega)$ and
$\chi_c(q,-\omega)=\chi_(q,\omega)$. At zero temperature
$S(q,\omega)=\chi_c(q,\omega)=\chi''(q,\omega)$ for
$\omega > 0$, therefore the sum rules given in Ref.
\onlinecite{Hohenberg} can be written as

\ba
m_1(q) &=& \int^{\infty}_{0}\frac{d\omega}{\pi} \frac{\chi''(q,\omega)}{\omega}
		= \frac{1}{2}\chi(q,\omega=0) \nonu
m_2(q) &=& \int^{\infty}_{0}\frac{d\omega}{\pi}\omega
		\frac{\chi''(q,\omega)}{\omega} =\chi_{c}(q,t=0) \nonu
m_3(q) &=& \int^{\infty}_{0}\frac{d\omega}{\pi} \omega^2
\frac{\chi''(q,\omega)}{\omega} =
-\frac{1}{2}\la [[H,n(q)],n(-q)] \ra \nonu
	&=& 2\la \cd_ic_{i+1}\ra(1-\cos(q))
\ea
\noin where $\chi(q,\omega=0)$ is the static susceptibility. These sum rules
can be easily generalized to higher moments:

\ba
m_l(q) &=& \int^{\infty}_0\frac{d\omega}{\pi} \omega^{n-1}
	\frac{\chi''(q,\omega)}{\omega} \nonu
    &=& \left \{ \begin{array}{rl} -\frac{1}{2} \la [[H,...,[H,n(q)]...],n(-q)]
	\ra &\hspace{0.1in} l \hspace{0.1in}$ even$ \\
	 \frac{1}{2} \la \{[H,...,[H,n(q)]...],n(-q)\} \ra
	 &\hspace{0.1in} l \hspace{0.1in}$ odd$ \end{array} \right. . \nonumber
\ea

\noin Apart from the first moments which is given by the static
susceptibility, all the other moments can be expressed as equal-time
correlation functions. Theoretically if all the moments are known,
one can obtain the $\chi''(q,t)$
and thus $\chi''(q,\omega)$. In real calculations, it is tedious to
calculate the commutators for higher moments, and there are more
and more new equal-time correlation functions appear in the expression of
higher moments.
However it is still reasonable to obtain the expression for the first several
moments using a symbolic manipulator, such as Mathematica, to calculate the
commutators. In this work we have calculated the
expressions for the first five moments. Details of the expressions for the
fourth and fifth moments will be given elsewhere \cite{pang}.

The second step is to obtain the moments by calculating those static
correlation functions with DMRG. The infinite lattice method
(see Ref. \onlinecite{White} for details) is used in our calculations
for open ended chains. $t=1$ is chosen, and states kept at each iteration
varies from 52 to 64.
We calculate the equal-time correlations, for example $\la n_i n_j \ra$, by
taking $j$ in the middle of the system. For a system which has parity and
translational symmetries, $\la n_j n_i \ra$ only depends on $|i-j|$. Therefore
$\la n_q n_{-q}\ra = \sum_l \la n_j n_{j+l}\ra e^{iql}$ is independent of $j$.
Since the calculations are done with open boundary condition, $\la
n_j n_{j+l} \ra$ depends on the position $j$. The boundary effect is larger
when $j$ or $j+l$ is closer to boundary, therefore
we choose $j$ at the center of the system. Also due to the open boundary,
the correlation $\la n_j n_{j+l} \ra$ has an even-odd oscillation in $j$.
We take the mean value of $\la n_j n_{j+l} \ra$ at $j$ even and odd,
which is close to the value with period boundary condition
for a system having the same size. When the system size goes to infinity,
the boundary effect can be neglected. We calculate the moments for
system sizes varying from 100 sites to 200 sites, and obtain their values
for infinite system by extrapolating the data.

The next step is to use MEM to obtain the dynamical correlation functions.
MEM has become standard way to extract maximum information from incomplete data
\cite{Brandt}. This method has been applied to the analytic
continuation of the Quantum Monte Carlo data
\cite{Silver}, and in this paper we apply a similar
method to extract the dynamic susceptibility $\chi''(q,\omega)$ from the
finite number of moments $m_l$ with the corresponding errors $\sigma_l$.
Defining $f(\omega)=\chi''(\omega)/\omega$, which is a positive definite,
as the distribution function, and  the entropy or the information function
$S=\sum_{\omega} f(\omega) - f(\omega) \log f(\omega)$.
By maximizing the entropy under the constrains
$m_l - \int_0^\infty \frac{d\omega}{\pi} \omega^{l-1} f(\omega) = 0$,
$f(\omega)$ has the following form
\be
f(\omega) = e^{- \sum_{l=1}^n(\lambda_l \omega^{l-1}) },
\ee
where $n$ is the number of moments and $\lambda_l$ are the Lagrange
multipliers. At this point one may try to find $\lambda_l$ by requiring the
$f(\omega)$ to satisfy the constrains without considering the error bars of
the moments. However, in general,
the error bars cannot be neglected. The kernel of the
transformation is singular, so small error bars in moments may produce
large errors in $f(\omega)$. By maximizing  the posterior probability
$e^{\alpha S - L}$ where $L \equiv \sum_l (m_l - \int_0^\infty
\frac{d\omega}{\pi} \omega^{l-1} f(\omega))^2/\sigma_l^2$, one can find
the most probable $f(\omega)$, which gives us the moments within
the range of error bars.

\begin{figure}[htb]
\centerline{\psfig{figure=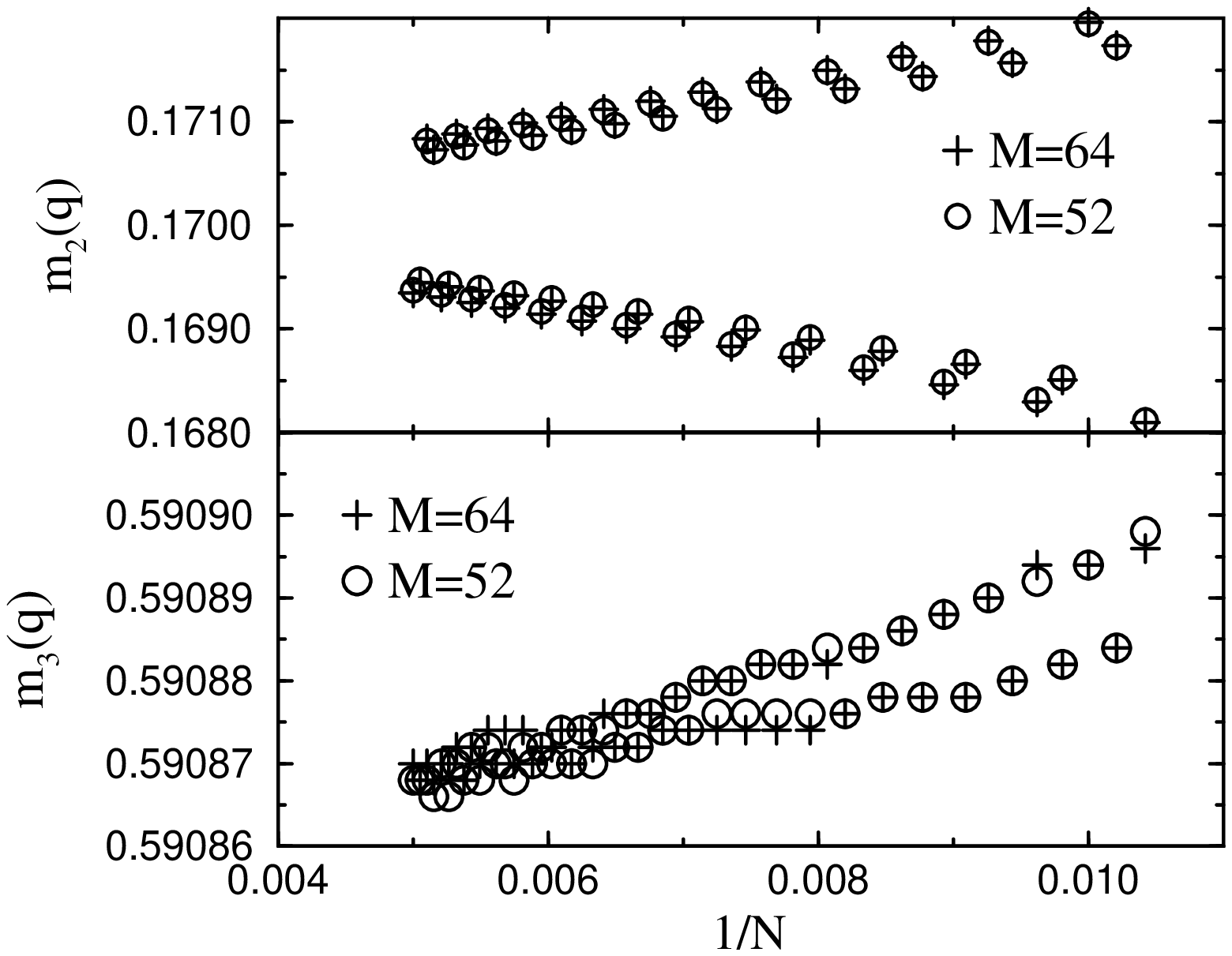,height=3.0in}}
\caption[]{Moments versus the inverse system size
$1/N\rightarrow 0$ for $V=2t$ and $q=\pi/2$, where $M$ is the number of
states kept at each iteration in DMRG calculations. The  extropolation is to
$1/N\rightarrow 0$, and the error is estimated by the different
extropolations caused by the errors in slope.
}
\label{extrapolate}
\end{figure}

Let us first discuss the extrapolation and the error bar of our DMRG results.
There are two major contributions to the error: that from finite size
effects and that from basis set transaction in the DMRG calculations.
The error bar of DMRG calculation for any finite size is obtained by
varying the number of states kept at each iteration, whereas the finite size
error is obtained by varying the system size. The asymptotic behavior
of correlation functions is known for this model \cite{Luther}, which
decay as a power of system size.
For a system with a gap, the extropolation should be done as an exponential
function of system size.
In Fig.~\ref{extrapolate},
we plot the second and third moments at $q=\pi/2$ for $V=2t$ as a
function of $1/N$, where $N$ is the number of sites of the system.
The error from basis set transaction produces the error in the extrapolating
values. We use this resultant error to estimate the error bar of the moments.
Extrapolating to $1/N \rightarrow 0$ gives $m_2 = 0.1700$,
and the error bar is estimated as $ 10^{-4}$. For the third moment
we have $m_3=0.59085$ and the estimated error bar is $2\times 10^{-5}$.
Actually the third moment is known exactly:
$m_3 =- \frac{2}{3} E (1-\cos(q))$ with the ground state energy per site
$E=-2(\ln2-1/4)$. The exact value at $\pi/2$ is 0.590863.

We test our method for the non-interacting case. In this case,
$\chi''(q,\omega)$ is known exactly:

\be
\chi''(q,\omega) = \frac{\theta(\omega -2t|\sin(q)|)
		\theta(4t|\sin(q/2)|-\omega)}{[16t^2\sin^2(q/2)-\omega^2]^{1/2}},
\ee

\begin{figure}
\centerline{\psfig{figure=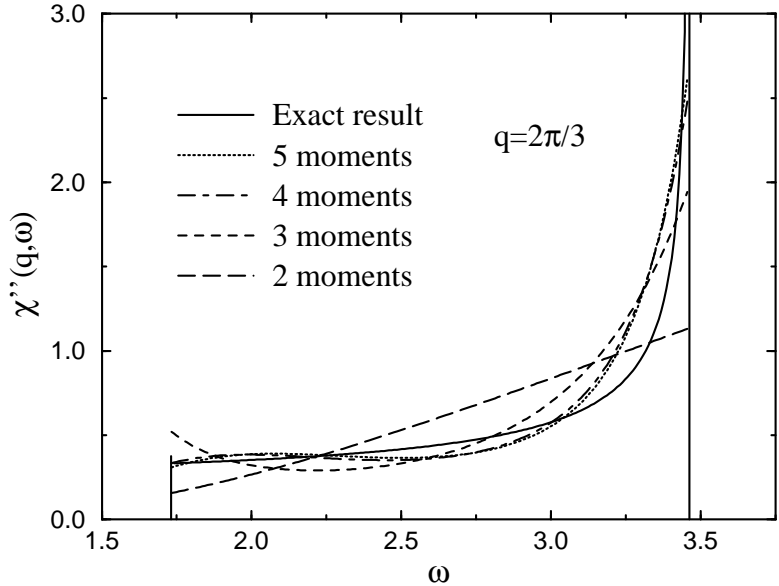,height=3.0in}}
\caption[]{The dynamical structure function $\chi''(q,\omega)$ for $V=0$
(the $XY$ model) and $q=2\pi/3$. We plot the results obtained by MEM
with different number of moments. Two solid vertical lines are the energy
boundaries.
}
\label{chi1}
\end{figure}

\noin where $\theta(x)$ is the step function. The moments can also be
calculated analytically. In Table \ref{table1}, we compare the moments
calculated by DMRG with the exact results. The error bars obtained by DMRG
provide a very good estimation.
Apart from the five moments, there are two more pieces
of information in this case: the energy boundaries $2t|\sin(q)| < \omega <
4t|\sin(q/2)|$ for $\chi''(q,\omega)$. Using the MEM, we obtain
$\chi(q,\omega)$ for $q=2\pi/3$. In Fig.~\ref{chi1},
we plot $\chi(q,\omega)$
obtained by MEM with different number of moments and the exact one from Eq.
(5). It shows that the $\chi(q,\omega)$ obtained by MEM
converge to the exact one when the number of moments is increased, and
$\chi(q,\omega)$ calculated with five moments is a good approximation
for the exact result. We have also calculated $\chi(q,\omega)$ for other $q$,
they have similar behavior.

\begin{figure}
\centerline{\psfig{figure=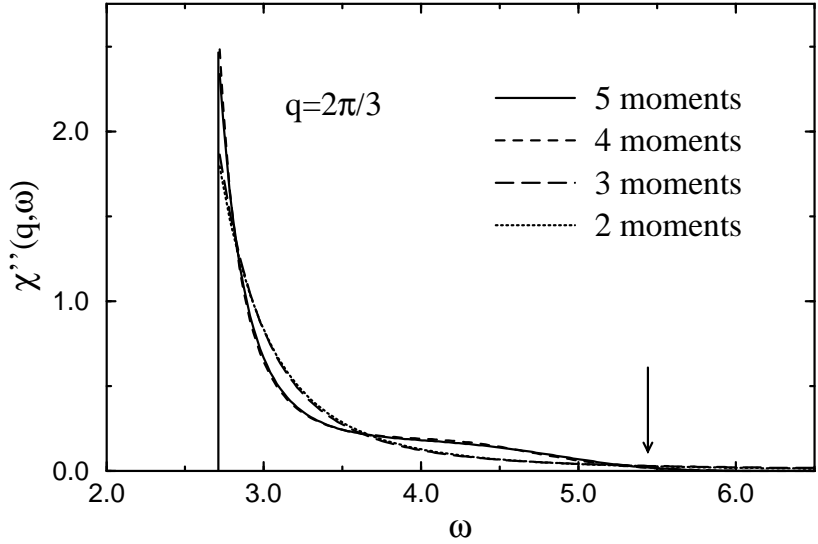,height=2.5in}}
\caption[]{The dynamical structure function $\chi''(q,\omega)$ for $V=2t$
(the Heisenberg model) and $q=2\pi/3$. We plot the results obtained by MEM
with different number of moments. The solid vertical line is the lower
boundary. The arrow marks the position of the upper
boundary for the two-spinon excited states.
}
\label{chi2}
\end{figure}

\begin{figure}
\centerline{\psfig{figure=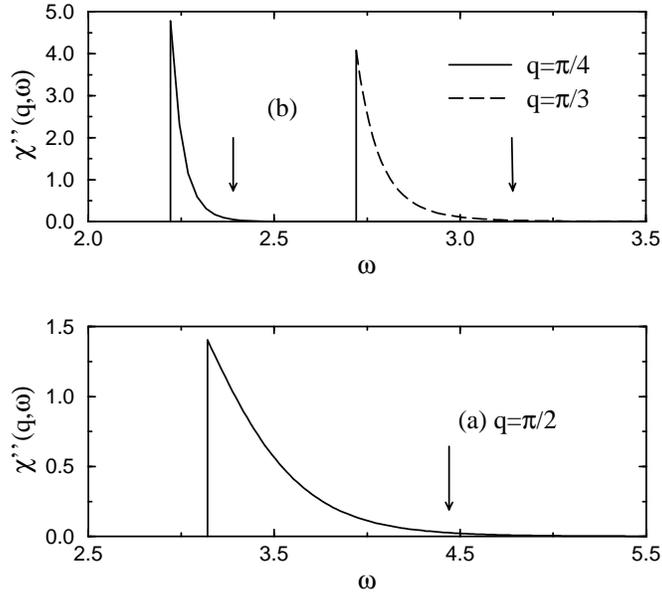,height=3.0in}}
\caption[]{The dynamical structure function $\chi''(q,\omega)$ for $V=2t$
(the Heisenberg model) at (a) $q=\pi/2$, and (b) $\pi/3$ and $\pi/4$.
$\chi''(q,\omega)$ are obtained by MEM with five moments. The solid vertical
lines are the lower boundaries for each momentum. The arrow marks the
position of the upper boundary of the two-spinon excited states.
}
\label{chi3}
\end{figure}

For the interacting case with $V=2t$, which corresponding to the Heisenberg
model,
the elementary excitations are known as $S=1/2$ objects \cite{Faddeev}
(spinons). The dispersion relation is
$\epsilon (q) = \pi t |\sin(q)|$ \cite{Cloizeaux1}, which
provides the lower bound of excitation energies for each momentum $q$.
The spectral weight is dominated by the continuum of the
two-spinon excited states \cite{Muller}, and the energy range for the
continuum is $t\pi|\sin(q)| < \omega < 2t|\pi\sin(q/2)|$.
Since the contributions from the
excited states of more than two spinons are finite, we only have
the low energy bound. In Fig.~\ref{chi2}, the
$\chi''(q,\omega)$ obtained by MEM with different number of moments are
plotted for $q=2\pi/3$. One can see the tendency of the curves as the
number of moments increase. $\chi''(q,\omega)$ tends to diverge at
lower-bound.
In Fig.~\ref{chi3}, $\chi''(q,\omega)$ is plotted for other momentums.
We marked the position of upper-bound for two-spinon excited states.
It is obvious that the contributions from excited states with more than
two spinons are finite, although they are small.

In conclusion, we have developed a numerical method for calculating the
ground-state dynamical correlation functions in one-dimensional quantum
systems based on the Density Matrix Renormalization Group Method and the
maximum entropy method. We demonstrate this method on the dynamical
density-density correlation $\chi''(q,\omega)$ of the spinless fermion system
with nearest neighbor interaction. For the non-interacting case, it
corresponds to the $XY$ model, and the dynamical density-density correlation
function obtained by our method shows a very good agreement with the exact
result. For the interacting case with $V=2t$, it corresponds to the
Heisenberg model, we obtain the $\chi''(q,\omega)$, which was not known
before. This method is a very general one, which can be applied to any
one-dimensional system with short range interaction like, e.g.
the Hubbard model, the $S=1$ Heisenberg model, the interacting fermion (or
boson) system with randomness.

We would like to acknowledge useful discussions with J.E. Gubernatis,
Shoudan Liang, and R.N.\ Silver. This work was
supported by the National Science Foundation grant No. DMR-9107563.
In addition MJ would like to acknowledge the support of the
NSF NYI program.

\def\baselinestretch{1.5}

\begin{table}
\caption[]{The comparison of the moments obtained by DMRG with the exact
results for $V=0$ and $q=2\pi/3$. In the DMRG calculations 64 states are kept
at each iteration, and the error bars are estimated by changing
the number of states kept and finite size scaling.}
\label{table1}
\begin{tabular}{llll}
   & EXACT & DMRG & ERROR \\
\tableline
$m_1(2\pi/3)$ & 0.121013 & 0.1211 & $1\times 10^{-3}$ \\
$m_2(2\pi/3)$ & 0.333333 & 0.33337   & $5\times 10^{-5}$ \\
$m_3(2\pi/3)$ & 0.954930 & 0.954928 & $5\times 10^{-6}$ \\
$m_4(2\pi/3)$ & 2.826993 & 2.8273 & $5\times 10^{-4}$ \\
$m_5(2\pi/3)$ & 8.594367 & 8.59434 & $5\times 10^{-5}$ \\
\end{tabular}
\end{table}

\end{document}